\documentclass[twocolumn,showpacs,superscriptaddress,aps,pra]{revtex4}
\usepackage{epsfig}
\usepackage{subfigure}
\usepackage{amssymb}
\usepackage{graphicx}
\usepackage{color}
\usepackage{amsfonts}
\usepackage{amscd}
\usepackage{hyperref}
\usepackage{amsmath}
\usepackage{epstopdf}

\usepackage{amsmath}
\usepackage{graphicx}
\usepackage{txfonts}

\begin{document}


\title{Robust  state preparation in quantum simulations of Dirac dynamics}

\author{Xue-Ke Song}
\affiliation{Department of Physical Chemistry, University of the Basque Country UPV/EHU, Apartado 644, 48080 Bilbao, Spain}
\affiliation{Department of Physics, Applied Optics Beijing Area Major Laboratory,
Beijing Normal University, Beijing 100875, China}

\author{Fu-Guo Deng}
\affiliation{Department of Physics, Applied Optics Beijing Area Major Laboratory,
Beijing Normal University, Beijing 100875, China}

\author{Lucas Lamata}
\affiliation{Department of Physical Chemistry, University of the Basque Country UPV/EHU, Apartado 644, 48080 Bilbao, Spain}

\author{J. G. Muga}
\affiliation{Department of Physical Chemistry, University of the Basque Country UPV/EHU, Apartado 644, 48080 Bilbao, Spain}

\date{\today }

\begin{abstract}
A non-relativistic system such as an ultracold trapped ion may perform a quantum
simulation of a Dirac equation dynamics under specific conditions.
The resulting Hamiltonian and dynamics are highly controllable, but the coupling between momentum and
internal levels poses some difficulties to manipulate the internal states accurately in wave packets.
We use invariants of motion to inverse engineer robust population inversion processes with a homogeneous,
time-dependent simulated electric field. This exemplifies the usefulness of
inverse-engineering techniques to improve the performance of
quantum simulation protocols.
\end{abstract}
\maketitle
%

%
%
%
%
%
%
\section{Introduction}
\label{sec1}
A recent highlight in the remarkable history of the Dirac equation
\cite{dirac,thaller} is the realization that non-relativistic systems such
as an ultracold trapped ion can obey this equation, with a proper
reinterpretation of symbols, under specific trapping conditions and
laser interactions \cite{Lamata,Gerritsma,Casanova,Gerri2}. In a one
dimensional setting (linear trap), two levels of the ion interacting
with laser fields set the basis that spans the relevant internal state subspace, whereas
orthogonal eigenvectors of the Dirac Hamiltonian with positive and
negative energies correspond to matter and antimatter solutions.
Similarly, different elements of the original Dirac equation, such
as the mass, or the constant playing the role of speed of light,
are mapped to atomic or interaction-dependent  properties. Different
interaction potentials may also be simulated, such as the ones for
homogeneous or linear electric fields \cite{Casanova}.
These
mappings and the controllability of trapped ions have been used to
observe experimentally simulations of relativistic
effects, like Zitterbewegung \cite{Gerritsma}, or Klein tunneling
\cite{Gerri2}. Trapped ions are in fact an example of a wider set of
non-relativistic ``Dirac systems'' that obey a Dirac dynamics,
for example in condensed matter \cite{Dirac1},  optics  \cite{Dirac2}, cold atoms
\cite{Salger,Zhang0}, or superconducting circuits \cite{Julen}.

The new physical platforms for Dirac dynamics are often easier to
manipulate than relativistic particles. In trapped ions, for
example, the effective (simulated) mass, speed of light, or electric
field may be changed in time.  This opens prospects for finding
and implementing new or exotic effects and carrying out further
fundamental studies. It also motivates a search for
manipulation protocols to achieve
specific goals \cite{deffner}. Shortcuts to adiabaticity (STA) \cite{Torrontegui}, a
group  of techniques to speed up adiabatic methods, possibly
following non-adiabatic routes, offer a suitable framework for the
task, and example cases have been worked out recently in the domain
of the Dirac equation \cite{deffner,Mugadirac}. STA are typically
highly flexible so that, apart from speeding up the processes, which
may be needed to avoid decoherence, the protocol may
satisfy further conditions, such as robustness with respect to noise
and/or systematic perturbations. Robust protocols have been
demonstrated for the Schr\"odinger equation  \cite{Ruschhaupt,Lu}, and,
as we shall see in this paper, can be extended as well to the Dirac
equation.

The study case we address here is a population inversion of the internal state, as a paradigmatic example of
single qubit operations,
making use of an effective time-dependent, homogeneous electric field.
Due to the structure of the Dirac Hamiltonian, a protocol designed to perform the inversion for a specific
momentum, say the average momentum of the wave packet,
in general will not work perfectly for other momenta. In other words, the momentum spread is a source
of systematic errors,
and our goal will be to design robust protocols with respect to momentum offsets
inherent in wave packets. The employment of inverse engineering and STA methods may enhance the toolbox of quantum simulations
and enable faster and more accurate protocols, which will presumably boost the field of quantum technologies.

The paper is organized as follows: In Sec. \ref{sec2} we set the model and Hamiltonian.
In Sec. \ref{sec3}
we give the solution via invariants. In Sec. \ref{sec4},
we put forward a robust invariant-based protocol to engineer the quantum state.
Sec. \ref{sec5} analyzes the robustness of the invariant-based shortcut protocols
against the systematic momentum error.
Sec. \ref{sec7} addresses a proposal to implement the robust protocol via a Dirac equation dynamics using trapped ions.
Finally Sec. \ref{sec6} summarizes and discusses the results.
\section{Driven Dirac dynamics with time-dependent vector field}
\label{sec2}
We focus now on a $1+1$-dimensional Dirac equation for a charged particle  moving in $x$-direction, which could
be simulated by ultra cold trapped ions and realizes
quantum relativistic effects \cite{Lamata,Gerritsma,Casanova}. It may be written as
\cite{deffner}
\begin{eqnarray}    
i\hbar|\dot{\Psi}(t)\rangle=[(-i\hbar c \partial_x+A(x,t))\sigma_x+mc^2\sigma_z]|\Psi(t)\rangle,
\label{electric}
\end{eqnarray}
where $|\Psi(t)\rangle$ is the two-component time-dependent wave function  for the particle with
mass $m$, the dot means time derivative, $c$ is the speed of
light, $\hbar$ is the Planck constant divided by $2\pi$, and
$\sigma_{x,y,z}$ are $2\times 2$ Pauli matrices in the basis
$\small{\left|1\right\rangle=\left(\begin{array}{c}
                                           1 \\
                                           0
                                         \end{array}
                                          \right)}$
and $\small{\left|2\right\rangle=\left(\begin{array}{c}
                                           0 \\
                                           1
                                         \end{array}
                                         \right)}$.
To implement  a time-dependent but spatially homogeneous electric
field, we set $A(x,t)$ as a purely time-dependent function,
$A(x,t)=\alpha_t$. Then the Hamiltonian reads
\begin{eqnarray}    
H=-i \hbar c \partial_x\sigma_x+\alpha_t\sigma_x+mc^2\sigma_z.
\label{hamiltonian}
\end{eqnarray}
Beware that $c$,  $m$, and the electric field must be reinterpreted in the simulated dynamics, as discussed in \cite{Lamata,Gerritsma,Casanova} and later in Sec. \ref{sec7}.
Note also that, whereas the two components of the state do not represent the spin in the relativistic interpretation \cite{thaller2},
the two levels $|1\rangle$ and $|2\rangle$ in the simulation simply become two bare internal levels of the ion.

Deffner \cite{deffner} used the fast-forward shortcut
technique \cite{ff1,ff2} to suppress ``production of pairs'' (transitions among positive and negative energy solutions)
in fast processes,
combining scalar and pseudoscalar potentials. Our goal here is instead
to induce a fast and robust population inversion among the bare levels. A different technique will be applied,
designing the time dependence
of the parameters in the Hamiltonian rather than adding terms to it. This is carried out by making use of invariants
of motion twice:
first to decompose the solution of the Dirac equation into independent subspaces for each plane wave,
and then, to describe and manipulate the solution
for the internal state amplitudes within each subspace \cite{Mugadirac}.
\section{Solutions via invariants} \label{sec3}
We shall find exact solutions of the Dirac equation in Eq.~(\ref{electric}) based on the Lewis and
Riesenfeld theory of invariants \cite{Lewis}.
For the Hamiltonian in Eq.~(\ref{hamiltonian}), let us assume that a nontrivial invariant exists with the form
\cite{Landim,Castro,Khantoul}
\begin{eqnarray}    
I=A(t)p+B(t)x+D(t),
\label{invariant}
\end{eqnarray}
where $A(t)$, $B(t)$, and $D(t)$ are $2\times 2$ matrices.
The invariant  should satisfy the equation
\begin{eqnarray}    
\frac{dI}{dt}=\frac{1}{i\hbar}[I,H]+\frac{\partial I}{\partial t}=0.
\label{dynamics}
\end{eqnarray}
Substituting Eqs.~(\ref{hamiltonian}) and (\ref{invariant})
into Eq.~(\ref{dynamics}) gives
\begin{eqnarray} 
[A,\sigma_{x}]&=&0,
\label{A}\\
{[B,\sigma_x]}&=&0,
\label{B}\\
\alpha_{t}[A,\sigma_{x}]+mc^{2}[A,\sigma_{z}]+c[D,\sigma_{x}]+i\hbar\dot{A}&=&0,\label{C}\\
\alpha_{t}[B,\sigma_{x}]+mc^{2}[B,\sigma_{z}]+i\hbar\dot{B}&=&0,\label{D}\\
icB\sigma_{x}+\alpha_{t}[D,\sigma_{x}]+mc^{2}[D,\sigma_{z}]+i\hbar\dot{D}&=&0.\label{E}
\end{eqnarray}
Expanding the matrices in the $su(2)$-basis, $A=a_{1}+a_{2}\sigma_{x}+a_{3}\sigma_{y}+a_{4}\sigma_{z}$
with $a_{i}$ an arbitrary real number for $i=1,2,3,4$, and
similarly for $B$ and $D$, the above equations are easy to solve.
From Eqs.~(\ref{A}) and (\ref{B}), we get
\begin{align} 
 A=a_{1}+a_{2}\sigma_{x}, \label{F}
 \\
 B=b_{1}+b_{2}\sigma_{x},\label{G}
\end{align}
where $a_{1}$, $a_{2}$, $b_{1}$, $b_{2}$ are to be determined.
Substituting Eq.~(\ref{F}) into Eq.~(\ref{C}), we have
\begin{eqnarray} 
 \dot{b}_{1}&=&\dot{b}_{2}=0,
 \\
 b_{2}&=&0.
\end{eqnarray}
Substituting Eq.~(\ref{G}) into Eq.~(\ref{D}), we have
\begin{eqnarray} 
 cd_{4}&=&mc^{2}a_{2},
\nonumber\\
d_{3}&=&0,
\nonumber\\
\dot{a}_{1}=\dot{a}_{2}&=&0.
\end{eqnarray}
Similarly, from Eq.~(\ref{E}), we find
\begin{eqnarray}  
cb_{1}+\hbar\dot{d}_{2}&=&0,
\nonumber\\
\alpha_{t}d_{4}&=&mc^{2}d_{2},
\nonumber\\
\dot{d}_{1}=\dot{d}_{4}&=&0.
\end{eqnarray}
The invariant can be then written as
\begin{eqnarray}    
I=(a_{1}p+b_{1}x+d_{1})+(a_{2}p+d_{2})\sigma_{x}+d_{4}\sigma_{z},
\end{eqnarray}
where $a_{1}$, $b_{1}$, $d_{1}$, and $d_{4}$ are constant.
If $\alpha_{t}$ is time-dependent, then
$d_{4}=d_{2}=0$, and therefore $b_{1}=a_{2}=0$. The invariant can be simplified
as
\begin{eqnarray}    
I=a_{1}p+d_{1}=a_{1}(p+\mathbb{C}), \label{I}
\end{eqnarray}
where $\mathbb{C}$ is a constant. This holds even for a time-dependent mass.
Consistently, the Heisenberg equations of motion for the system Eq.~(\ref{hamiltonian})
are
\begin{eqnarray}    
\frac{dp}{dt}=0,\;\; \frac{dx}{dt}=c\sigma_x.
\end{eqnarray}
In other words, the momentum operator is  invariant, which may be interpreted as  the initial momentum $p_{0}$ \cite{zhang}, as shown below making use of a different frame.

The solutions of the time-dependent Dirac equation may be written as linear superpositions
of eigenvectors of the invariant \cite{Lewis}.
Since the eigenfunctions of the invariant take the plane-wave form
$e^{ip_{0}x/\hbar}$ with $p_{0}$ a real number,
we assume the existence of plane wave solutions of Eq. (\ref{hamiltonian}) according to the ansatz
\begin{eqnarray}    
|\phi(t)\rangle=e^{ip_{0}x/\hbar}|\phi_{p_{0}}(t)\rangle, \label{wavepacket}
\end{eqnarray}
where $|\phi_{p_{0}}(t)\rangle$ is a $2\times1$ vector that depends on the parameters
$p_{0}$ and $t$.

Substituting Eq.~(\ref{wavepacket})
into the time-dependent Dirac equation in Eq.~(\ref{electric}) gives the following
reduced ($2\times2$) Dirac equation for the vector $|\phi_{p_{0}}(t)\rangle$,
\begin{eqnarray} 
i\hbar|\dot{\phi}_{p_{0}}(t)\rangle=H_{p_{0}}|\phi_{p_{0}} (t)\rangle, \label{new}
\end{eqnarray}
where
\begin{eqnarray} 
H_{p_{0}}=cp_{0}\sigma_{x}+\alpha_{t}\sigma_{x}+mc^{2}\sigma_{z}.
\label{hami0}
\end{eqnarray}
By superposing plane wave solutions, general (wave packet) solutions are found, of the form
\begin{equation} 
|\Psi(t)\rangle=\int_{-\infty}^{\infty}a{(p_{0})}|\phi_{p_0} (t)\rangle dp_{0},
\end{equation}
where each (momentum) component evolves with its own $2\times 2$ Hamiltonian $H_{p_0}$,
so that the corresponding global (wave packet) populations for $|1 \rangle$ and $|2 \rangle$ are given by
\begin{eqnarray}    
P_{k}=\int_{-\infty}^{\infty}|a(p_{0})|^2P_{k}(p_0)dp_{0},
\label{gau}
\end{eqnarray}
where $k=1,2$ and $P_{k}(p_0)=|\langle k|\phi_{p_{0}}(t)\rangle|^{2}$ $(k=1,2)$ are the populations for each
momentum
in the basis $\{|1\rangle, |2\rangle\}$. In the numerical examples we take a Gaussian
function $|a(p_{0})|^2=\frac{1}{\sqrt{2\pi}\sigma}\exp(-p_{0}^{2}/\sigma^{2})$.

The homogeneous electric field is more often represented by a linear scalar potential.
To find this representation and see the equivalence with our treatment,  we change the frame
by  means of the  unitary transformation
$U=e^{-i\alpha_{t}x/(\hbar c)}$. The effective Hamiltonian becomes
\begin{eqnarray}    
H_{u}=U^{\dagger}HU-i\hbar U^{\dagger}\dot{U}=cp\sigma_{x}+mc^{2}\sigma_{z}-\dot{\alpha}_{t}x/c,
\label{HU}
\end{eqnarray}
where we have used the Hausdorff expansion, which can be truncated here exactly, as $e^{\xi x}He^{-\xi x}=H+\xi[x,H]$,
with $\xi=-i\alpha_{t}/(\hbar c)$. The homogeneous field is now represented by a linear scalar potential of time-varying slope.
The plane wave solutions transform as
$|\phi_{u}(t)\rangle=U^{\dagger}|\phi(t)\rangle=e^{i(p_{0}+\alpha_{t}/c)x/\hbar}|\phi_{p_{0}}(t)\rangle$
so they get a time-dependent momentum
and the invariant of $H_u$ becomes (as it may be seen by repeating the steps after Eq. (\ref{dynamics})
for $H_u$) $I_{u}=\mathbb{C}(p-\alpha_{t}/c)$.
Since the two frames are unitarily connected, in what follows we shall use for simplicity
the one  based on $H$.
\section{Robust quantum state engineering} \label{sec4}
\subsection{Invariant-based shortcuts to adiabaticity for driven Dirac dynamics}
\label{sec41}
The Hamiltonian $H_{p_{0}}$ in (\ref{hami0}) for the Dirac system with spatially homogeneous
electric field reads in matrix form
\begin{eqnarray} 
H_{p_{0}}& = \left(
              \begin{array}{ccc}
                mc^{2}                           & cp_{0}+\alpha_{t}        \\
               cp_{0}+\alpha_{t}                              &-mc^{2}      \\
              \end{array}
            \right). \label{hp0}
\end{eqnarray}
If the functions of time $m(t)$ and $\alpha_{t}$ are given,
different values of $p_{0}$ imply  different $2\times2$ Hamiltonians,
with different solutions of the Dirac equation (\ref{new}). If we design  $m(t)$ and $\alpha_{t}$
by inverse engineering so as to induce a population inversion (or some other operation), say at $p_0=0$,
which we assume to be the average momentum of a wave packet,
the solution for any other momentum will generally fail to satisfy the intended task.
In other words, the spread of $p_{0}$ in a wave packet can affect the dynamics
and induce errors. Therefore, it is necessary
to design protocols robust with respect to  the momentum spread. The
perturbed Hamiltonian $H_{p_{0}}$ can be decomposed as
$H_{p_{0}}=H_{0}(t)+H_{1}(t)$,
where
$\small{H_{0}(t)= \left(
              \begin{array}{ccc}
                mc^{2}                           & \alpha_{t}        \\
               \alpha_{t}                        &-mc^{2}      \\
              \end{array}
            \right)}$
is the unperturbed Hamiltonian and
$\small{H_{1}(t)=c \left(
              \begin{array}{ccc}
               0                          & p_{0}        \\
               p_{0}                        &0      \\
              \end{array}
            \right)}$
is the ``systematic error'' Hamiltonian. In the following, adopting the
standard notation for two-level Hamiltonians in quantum optics, $\frac{\hbar}{2}\Delta(t)=mc^{2}$ and
$\frac{\hbar}{2}\Omega(t)=\alpha_{t}$, in terms of a detuning $\Delta$, and a Rabi frequency $\Omega$,
we write
\begin{eqnarray} 
H_{0}(t)& =\frac{\hbar}{2} \left(
              \begin{array}{ccc}
                \Delta                          & \Omega       \\
              \Omega                 &-\Delta      \\

              \end{array}
            \right). \label{qh}
\end{eqnarray}
The instantaneous adiabatic eigenstates of $H_0(t)$ are
\begin{eqnarray} 
| E_{+}(t)\rangle&=&\cos\left(\frac{\varphi}{2}\right)|1 \rangle+\sin\left(\frac{\varphi}{2}\right)|2 \rangle,
\\
| E_{-}(t)\rangle&=&\sin\left(\frac{\varphi}{2}\right)|1 \rangle-\cos\left(\frac{\varphi}{2}\right)|2 \rangle,
\end{eqnarray}
with the mixing angle $\varphi=\arctan(\Delta/\Omega)$ and
the corresponding adiabatic energies $E_{\pm}(t)=\pm\frac{\hbar}{2}\sqrt{\Delta^{2}+\Omega^{2}}$.

For  this time-dependent
$2\times 2$ Hamiltonian $H_{0}$, there exists a dynamical invariant $I_{0}$,
not to be confused with the momentum invariant of Eq. (\ref{hamiltonian}).
This  invariant in the internal-state subspace can be written as
\cite{Lewis,Muga,Chen,Torrontegui}
\begin{eqnarray} 
I_{0}(t)& =\frac{\hbar}{2}\Omega_{0} \left(
              \begin{array}{ccc}
               \cos\theta                          & \sin\theta e^{i\beta}       \\
              \sin\theta e^{-i\beta}                 &-\cos\theta      \\
              \end{array}
            \right),\label{io}
\end{eqnarray}
where $\Omega_{0}$ is an arbitrary constant (angular) frequency to keep
$I_{0}(t)$ with dimensions of energy, and $\theta$ and $\beta$ are auxiliary time-dependent angles.
Using  Eqs.~(\ref{qh}) and (\ref{io})
in Eq.~(\ref{dynamics}) we find the  differential equations
\begin{eqnarray} 
\dot{\theta}&=&\Omega\sin\beta,
\label{omega}
\\
\dot{\beta}&=&\Omega\cot\theta\cos\beta-\Delta.
\label{beta}
\end{eqnarray}
The eigenstates of the invariant are
\begin{eqnarray}  
|\phi_{+}(t)\rangle &=&\left(\begin{array}{c}
                                                      \cos\left(\theta/2\right)e^{i\beta/2} \\
                                                      \sin\left(\theta/2\right)e^{-i\beta/2}
                                                    \end{array}\right),
\label{inveigen+}
\\
|\phi_{-}(t)\rangle &=&\left(\begin{array}{c}
                                                     \sin\left(\theta/2\right)e^{i\beta/2} \\
                                                      -\cos\left(\theta/2\right)e^{-i\beta/2}
                                                    \end{array}\right),
\label{inveigen-}
\end{eqnarray}
which satisfy
$I_{0}|\phi_{n}(t)\rangle=\lambda_{n}|\phi_{n}(t)\rangle$
$(n=\pm)$ with the eigenvalues $\lambda_{\pm}=\pm\hbar\Omega_{0}/2$.
The general solution of the time-dependent Schr\"{o}dinger equation, according
to the theory of Lewis and Riesenfeld \cite{Lewis}, can be written as a linear combination
$|\Phi_s\rangle=\sum_{n=\pm}c_{n}e^{i\epsilon_{n}}|\phi_{n}\rangle$,
where $c_{\pm}$ are time-independent amplitudes, and the $\epsilon_{\pm}$
are the Lewis-Riesenfeld phases
\begin{eqnarray}    
\epsilon_{\pm}(t)=\frac{1}{\hbar}\int_{0}^t\left\langle\phi_{\pm}(t')\left|i\hbar\frac{\partial}{\partial t'}
-H_{0}(t')\right|\phi_{\pm}(t')\right\rangle dt'.  \label{LR}
\end{eqnarray}
Then, two orthogonal solutions can be constructed as
\begin{eqnarray} 
|\psi_{0}(t)\rangle=e^{-i\gamma (t)/2}|\phi_{+}(t)\rangle
=e^{-i\gamma (t)/2}\left(\begin{array}{c}
                                                \cos(\theta/2)e^{i\beta/2} \\
                                                      \sin(\theta/2)e^{-i\beta/2}
                                                    \end{array}\right),
\label{psi0}
\end{eqnarray}
and
\begin{eqnarray} 
|\psi_{\perp}(t)\rangle=e^{i\gamma (t)/2}|\phi_{-}(t)\rangle
=e^{i\gamma (t)/2}\left(\begin{array}{c}
                                                      \sin(\theta/2)e^{i\beta/2} \\
                                                      -\cos(\theta/2)e^{-i\beta/2}
                                                    \end{array}\right),
\end{eqnarray}
where $\gamma=2\epsilon_{-}=-2\epsilon_{+}$ and $\langle\psi_{0}(t)|\psi_{\perp}(t)\rangle=0$
for all times. Thus, by using Eqs.~(\ref{omega}) and (\ref{LR}), we find
\begin{eqnarray}    
\dot{\gamma}=\frac{\Omega\cos\beta}{\sin\theta}=\frac{\dot{\theta}\cos\beta}{\sin\theta\sin\beta}. \label{gamma}
\end{eqnarray}
Our aim is to design invariant-based shortcuts to achieve a
population inversion from state $|1\rangle$ to state $|2\rangle$, up to a
global phase factor, along the invariant eigenstate
$|\phi_{+}(t)\rangle$ in a given time $t_{f}$.
We therefore write down the boundary conditions for $\theta$ to
guarantee the desired initial and final states,
\begin{eqnarray}    
\theta(0)=0,\;\;\;\;\;\;\;\theta(t_{f})=\pi.
\label{BC1}
\end{eqnarray}
In addition, if
we impose $[H_{0}(0),I_{0}(0)]=0$ and
$[H_{0}(t_{f}),I_{0}(t_{f})]=0$ so that the Hamiltonian $H_{0}(t)$
and the invariant $I_{0}(t)$ share common eigenstates
at initial and final times, we have the following additional boundary conditions,
\begin{eqnarray}    
\Omega(0)&=&0,\;\;\;\;\;\;\;\dot{\theta}(0)=0,
\nonumber \\
\Omega(t_{f})&=&0,\;\;\;\;\;\;\;\dot{\theta}(t_{f})=0. \label{BC2}
\end{eqnarray}
The Rabi frequency and detuning
leading to a fast population inversion are  determined from Eqs.~(\ref{omega}) and (\ref{beta}),
choosing a convenient function of $\beta$, and interpolating $\theta$ to satisfy the
boundary conditions (\ref{BC1}) and (\ref{BC2}).
\subsection{Robust shortcuts against systematic momentum errors}
\label{sec42}
%
%
\begin{figure}[!h]
\begin{center}
\includegraphics[width=7 cm,angle=0]{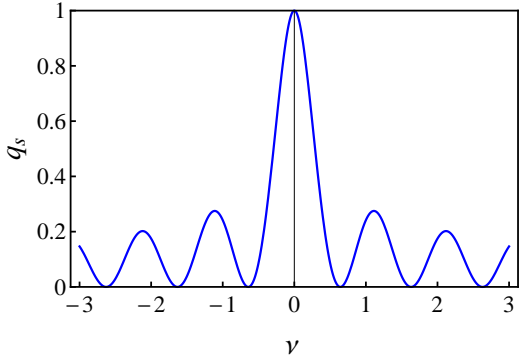}
\caption{(Color online) Systematic error sensitivity $q_{s}$ in Eq. (\ref{47}). As in all figures we use dimensionless units
with $c=\hbar=t_f=1$.
For specific values of $|\nu|$,  $q_{s}=0$ is satisfied, in particular at the minimal value $|\nu|=0.643$.}
\label{fig1}
\end{center}
\end{figure}

\begin{figure}[!h]
\begin{center}
\includegraphics[width=8.0 cm,angle=0]{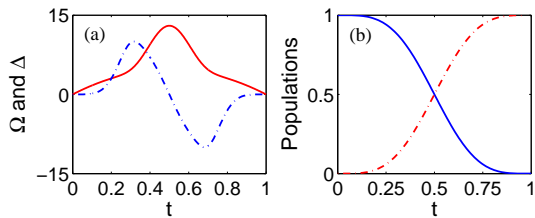}
\caption{(Color online) (a) The Rabi frequency $\Omega$ (red, solid line) and
detuning $\Delta$ (blue, dotted-dashed line) in our optimal protocol.
(b) Time evolution of the populations $P_{1}(0)$ (blue, solid line)
and $P_{2}(0)$ (red, dotted-dashed line) during the population inversion.
We have used $\nu=0.643$ and $p_0=0$. }
\label{fig2}
\end{center}
\end{figure}

\begin{figure}[!h]
\begin{center}
\includegraphics[width=6 cm,angle=0]{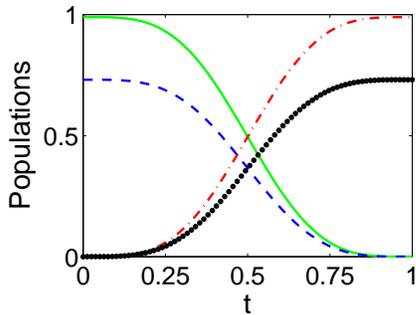}
\caption{(Color online) Time evolution of the populations $P_{1}$ of a Gaussian wave packet centered at
zero momentum (green, solid line
and blue, dot-dashed line for $\sigma=0.3$ and $\sigma=0.9$, respectively)
and $P_{2}$ (red, dotted-dashed line
and black circles for $\sigma=0.3$ and $\sigma=0.9$, respectively) by averaging over all
momenta $p_{0}$, see Eq. (\ref{gau}), during the population inversion.
$H_0$ as in Fig. \ref{fig2} (a). Compare to the result for a plane wave, $p_0=0$, in Fig. \ref{fig2} (b).}
\label{fig3}
\end{center}
\end{figure}

To construct invariant-based shortcuts robust against the systematic
momentum errors, we use perturbation
theory up to $\mathcal{O}(p_{0}^{2})$ to find the time
evolution of the quantum state governed by $H_{p_{0}}$ that starts as $|\psi_0(0)\rangle$,
\begin{eqnarray} 
&&\hspace{-1cm}|\psi(t_{f})\rangle
= |\psi_{0}(t_{f})\rangle -\frac{i}{\hbar}\int_{0}^{t_{f}}dt\hat{U}_{0}(t_{f},t)H_{1}(t)|\psi_{0}(t)\rangle
\nonumber\\
&&\hspace{-1cm}-\frac{1}{\hbar^{2}}\!\int_{0}^{t_{f}}\!\!\!dt\!\!\!\int_{0}^{t}\!\!\!dt'\hat{U}_{0}(t_{f},t)H_{1}(t)\hat{U}_{0}(t,t')H_{1}(t')|\psi_{0}(t')\rangle
+\cdot\cdot\cdot,
\end{eqnarray}
where $|\psi_{0}(t)\rangle$ is the unperturbed solution and $\hat{U}_{0}(s,t)=|\psi_{0}(s)\rangle \langle \psi_{0}(t)|
+|\psi_{\perp}(s)\rangle \langle \psi_{\perp}(t)|$
is the unperturbed time evolution operator. We assume that the error-free
$(p_{0}=0)$ scheme works perfectly, i.e., $|\psi_{0}(0)\rangle =|1\rangle$, $|\psi_{0}(t_{f})\rangle =|2\rangle$.
Then, the probability of the excited state at the final time for
$t_{f}$ and momentum $p_0$ is
\begin{eqnarray}    
\!\!P_{2}(p_0)\!=\!|\langle \psi_{0}(t_{f})|\psi(t_{f})\rangle |^2\!\!=\!\!
1\!\!-\!\!\frac{1}{\hbar^{2}}\Bigg|\!\int_{0}^{t_{f}}\!\!\!dt\langle \psi_{\perp}(t)|H_{1}(t)|\psi_{0}(t)\rangle \Bigg|^{2}\!.
\end{eqnarray}
Defining the systematic error sensitivity as \cite{Ruschhaupt,Lu}
\begin{eqnarray}    
q_{s}\coloneqq-\frac{1}{2}  \frac{\partial^{2}P_{2}(p_0)}{\partial p_{0}^{2}}\Bigg|_{p_{0}=0}=
-\frac{\partial P_{2}(p_0)}{\partial(p_{0}^{2})}\Bigg|_{p_{0}=0},
\end{eqnarray}
we have
\begin{eqnarray}    
q_{s}=\frac{c^{2}}{\hbar^{2}}\left|\int_{0}^{t_{f}}dte^{-i\gamma}(-i\sin\beta-\cos\theta\cos\beta)\right|^{2}.
\end{eqnarray}
For a flat $\pi$ pulse,  $\beta=\pi/2$,
and $\theta=\pi t/t_{f}$, so
$\dot{\theta}=\pi/t_{f}$, $\Omega=\pi/t_{f}$, $\Delta=0$,
and $\dot{\gamma}=0$. This  gives
\begin{eqnarray}    
q_{s}(\pi\, {\rm pulse})=
\frac{c^{2}t_{f}^{2}}{\hbar^{2}}.
\label{43}
\end{eqnarray}
Optimally robust  invariant-based shortcuts are now defined as those that make the systematic error
sensitivity zero. Following \cite{Daems}, we could try the simple Fourier series type of ansatz
\begin{eqnarray}    
\gamma=2\theta+\nu\sin(2\theta),
\end{eqnarray}
where $\nu$ is a real number that may be varied to nullify $q_{s}$.
(It is possible to extended this ansatz to make further derivatives zero as in \cite{Daems}.)
Alternatively we use \cite{Ruschhaupt}
\begin{eqnarray}    
\gamma=\nu[2\theta-\sin(2\theta)].
\label{gamma2}
\end{eqnarray}
 Both ansatzes are valid and nullify $q_s$ for different values of $\nu$.
 They lead  approximately to the same pulse area $A=\int_{0}^{t_{f}}\Omega(t)dt$,
 but the second one provides simpler expressions of
 $\beta$, $\Omega$ and $\Delta$, using  Eqs.~(\ref{omega}), (\ref{beta}), and ~(\ref{gamma}),
 so it is preferred here.
Specifically, using Eqs. (\ref{gamma}) and (\ref{gamma2}), the parameter $\beta$ takes the form
\begin{eqnarray}    
\beta={\rm{arccot}}(4\nu\sin^{3}\theta).
\end{eqnarray}
This gives $\beta(0)=\beta(t_f)=\pi/2$ so that the invariant eigenstate $|\phi_+(t)\rangle$, see Eq. (\ref{inveigen+}),
evolves from $|1\rangle$ to $|2\rangle$ up to phase factors, $|\phi_+(0)\rangle=e^{i\pi/4}|1\rangle$ and $|\phi_+(t_f)\rangle=
e^{-i\pi/4}|2\rangle$.
Finally, the systematic errors sensitivity is given by
\begin{eqnarray}    
q_{s}=\frac{c^{2}}{\hbar^{2}}\left|\int_{0}^{t_{f}}dte^{-i\nu[2\theta-
\sin(2\theta)]}\frac{-i-4\nu\sin^{3}\theta\cos\theta}{\sqrt{1+16\nu^{2}\sin^{6}\theta}}\right|^{2}.
\label{47}
\end{eqnarray}
Fig. \ref{fig1} shows the
systematic error sensitivity versus $\nu$, passing through
zeroes of $q_{s}$. (In all numerical calculations we use dimensionless units with $c=\hbar=t_f=1$. The dimensionless
effective mass generally depends on time so it is not made one as usual.)
The corresponding Rabi frequency and detuning are
\begin{eqnarray} 
\Omega&=&\dot{\theta}\sqrt{1+16\nu^{2}\sin^{6}\theta},\\
\Delta&=&16\nu\sin^{2}\theta\cos\theta\,\dot{\theta}\frac{1+4\nu^{2}\sin^{6}\theta}{1+16\nu^{2}\sin^{6}\theta}.
\end{eqnarray}
$\Omega$ increases monotonously with $\nu$ so we choose the smaller value consistent with $q_s=0$,
$\nu_{m}=0.643$, to minimize $\Omega$ along the evolution path. In addition, to interpolate at
intermediate times, we assume a polynomial ansatz $\theta=\sum_{j=0}^3a_{j}t^{j}$, where the
coefficients $a_{j}$ are found  by solving the equations set by the boundary
conditions on  $\theta$ and its derivative, see Eqs. (\ref{BC1}) and (\ref{BC2}). The time-dependent
$\Omega$ and $\Delta$ are shown in Fig. \ref{fig2} (a), with absolute value maxima $|\Omega_{m}|\simeq13$ and $|\Delta_{m}|\simeq10$.
For the specified $H_{0}(t)$ in Eq.~(\ref{qh}), corresponding to $p_0=0$, we solve
$H_{0}|\phi_{0} (t)\rangle=i\hbar|\dot{\phi_{0}} (t)\rangle$
numerically by a Runge-Kutta method with an adaptive step, and get the
time evolution of the populations $P_{k}(p_0=0)$ for the optimal protocol represented in Fig. \ref{fig2} (a).
Fig. \ref{fig2} (b) shows the population inversion between $|1 \rangle$ and $|2 \rangle$.
By contrast, solving the dynamics separately for each $p_0$ with $H_{p_0}$, and averaging
the populations $P_k(p_0)$ according to Eq. (\ref{gau}),  Fig. \ref{fig3} shows the change of the global population $P_{k}$
for  Gaussian wave packets with  $\sigma=0.3$ and
$\sigma=0.9$, respectively. The population inversion is still accurate for $\sigma=0.3$, but
by further increasing the momentum width, it eventually must fail.
$P_2(p_0)$ is shown in the next section, making explicit the momentum-width window
where a perfect inversion can be achieved.

\begin{figure}[!h]
\begin{center}
\includegraphics[width=8.0 cm,angle=0]{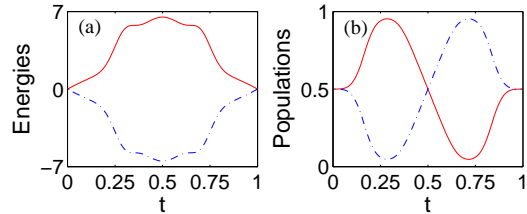}
\caption{(Color online) (a) The adiabatic energies of Hamiltonian $H_{0}(t)$:
$E_{+}(t)$ (red, solid line) and $E_{-}(t)$ (blue, dotted-dashed line).
(b) The adiabatic time evolution of the populations of level $|1\rangle$ for the positive (red, solid line)
and negative (blue, dotted-dashed line) energy eigenstates of Hamiltonian $H_{0}(t)$.
$\Omega$ and $\Delta$ are as in Fig. \ref{fig2} (a).}
\label{fig4}
\end{center}
\end{figure}

\begin{figure}[!h]
\begin{center}
\includegraphics[width=6.0 cm,angle=0]{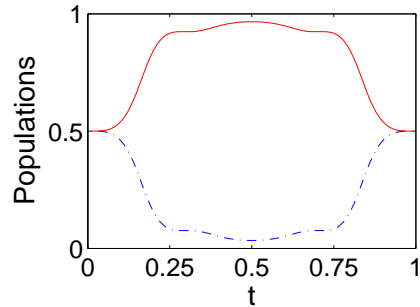}
\caption{(Color online) Populations of energy eigenstates along the invariant eigenstate $|\phi_+(t)\rangle$,
$|\langle E_+(t)|\phi_+(t)\rangle|^2$ (red, solid line)
and $|\langle E_-(t)|\phi_+(t)\rangle|^2$  (blue, dotted-dashed line), for the  optimal $\Omega(t)$ and
$\Delta(t)$  in Fig. \ref{fig2} (a).}
\label{fig5}
\end{center}
\end{figure}

We plot the adiabatic (instantaneous) eigenenergies of $H_{0}(t)$  in Fig. \ref{fig4} (a) for the optimal protocol.
Note the degeneracy at the edge times due to the vanishing of $\Delta$ and $\Omega$.
Fig. \ref{fig4} (b) depicts the adiabatic time evolution of the populations of level $|1\rangle$ in both eigenstates,
$|\langle1|E_{+}(t)\rangle|^{2}$ and $|\langle1|E_{-}(t)\rangle|^{2}$. In addition, Fig. \ref{fig5} depicts the instantaneous
populations of positive and negative energy  eigenstates for the invariant eigenstates,
$|\langle E_{+}(t)|\phi_{+}(t)\rangle|^{2}$ and $|\langle E_{-}(t)|\phi_{+}(t)\rangle|^{2}$. While the positive energy solution
dominates most of the time, both are equally important at boundary times.

%
%
%
%
%
\section{Robustness against wave packet momentum spread}
\label{sec5}

\begin{figure}[!h]
\begin{center}
\includegraphics[width=7.2 cm,angle=0]{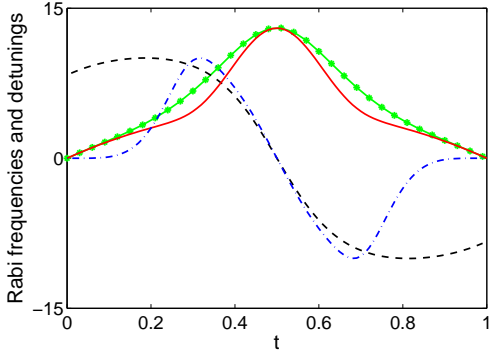}
\caption{The Rabi frequency $\Omega_{s}$ (green, dotted-star line), and detuning
and $\Delta_{s}$ (black, dashed line) are determined by Eqs. (\ref{omega}) and (\ref{beta})
with angles $\theta_s(t)=\sum_{j=0}^3a_{j}t^{j}$ and $\beta_s(t)=\sum_{j=0}^4b_{j}t^{j}$ in
simple invariant-based shortcuts, together with ``optimal'' $\Omega(t)$ (red, solid line) and
$\Delta(t)$ (blue, dotted-dashed line)  in Fig. \ref{fig2} (a).}
\label{fig6}
\end{center}
\end{figure}

\begin{figure}[!h]
\begin{center}
\includegraphics[width=7.0 cm,angle=0]{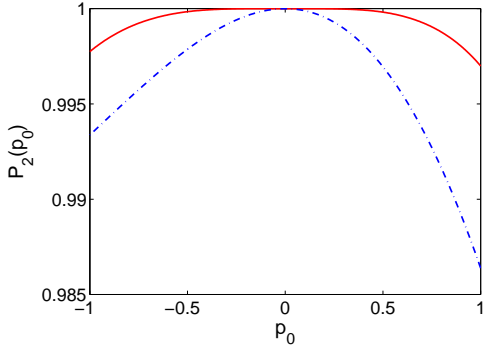}
\caption{Probability $P_{2}(p_0)$ at the final time $t_{f}=1$
versus systematic momentum noise $p_{0}$ by solving numerically Eq. (\ref{new})
with the Hamiltonian (\ref{hp0}) based on the optimal
invariant-based shortcut of Fig. \ref{fig2} (a) (zero sensitivity, red, solid line), and simple ones (nonzero sensitivity, blue,
dotted-dashed line).}
\label{fig7}
\end{center}
\end{figure}

We now test the stability of the optimal invariant-based protocol of the previous section
with respect to the momentum spread in wave packets, compared to a simple invariant-based shortcut for which the sensitivity
is not zero. Both protocols should invert the population along the invariant
eigenstate $|\phi_{+}(t)\rangle$ in a given time $t_{f}$ for $p_0=0$.
Let us denote by a subscript ``s" the auxiliary angles $\theta_s(t)$ and $\beta_s(t)$
and the Hamiltonian functions $\Omega_{s}$, $\Delta_{s}$ for the simple protocol with nonzero sensitivity.
To perform a fair comparison, we impose the same maxima of Rabi frequency
and detuning for the two protocols. We also take $\theta_s(t)=\theta(t)$ and $\beta_s(0)=\beta_s(t_f)=\pi/2$ for simplicity.
Setting $\beta_s(t_{f}/2)=2\pi/17$
the maximum of the Rabi frequency becomes $\Omega^m_{s}\simeq13$,
as in the optimal invariant-based shortcut. Moreover, the derivative of $\beta_s(t)$
at boundary times is chosen as $\dot{\beta}_s(0)=-\dot{\beta}_s(t_{f})=-15\pi/(17t_{f})$, so that the maximal
detuning $|\Delta^m_{s}|\simeq10$ at initial and final times is the same as for the optimal protocol.
$\beta_s(t)$ is interpolated at intermediate times with a polynomial
ansatz $\beta_s(t)=\sum_{j=0}^4b_{j}t^{j}$, where the coefficients $b_{j}$ are found by solving the boundary conditions.
With the determined $\beta_s(t)$ and $\theta_s(t)$, the Rabi frequency $\Omega_{s}(t)$ and
detuning $\Delta_{s}(t)$ in the simple invariant-based shortcut can be
calculated from Eqs. (\ref{omega}) and (\ref{beta}). They are plotted in Fig. \ref{fig6}, together with the Rabi frequency
and detuning of the optimal protocol of Fig. \ref{fig2} (a), which in fact has a slightly smaller pulse area.
By making use of Eq. (\ref{hp0}) with $mc^{2}=\frac{\hbar}{2}\Delta(t)$ and
$\alpha_{t}=\frac{\hbar}{2}\Omega(t)$ to solve numerically Eq. (\ref{new})
with the initial state $|1\rangle$,
the excitation probabilities $P_{2}(p_0)$ at final time $t_{f}=1$
based on the different invariant-based shortcuts are depicted in
Fig. \ref{fig7}, which demonstrates the robustness of the optimal protocol.
If needed, it is possible to systematically increase the width of the plateau as in
\cite{Daems}, by nullifying higher derivatives of the population at $p_0=0$.
\section{Trapped-ion implementation}
\label{sec7}
Even though the basic structure of a trapped-ion implementation of a 1+1 Dirac equation
was already proposed in Refs.~\cite{Lamata,Gerritsma,Casanova}, in our current formalism the simulated mass and electric field should be time-dependent and highly controllable, which is a novelty with respect to previous Dirac equation proposals and experiments in trapped ions. The high degree of laser control in trapped ions enables this kind of approach, given that laser amplitudes can be turned on and off in situ and their profiles designed according to the requirements of the proposed protocol.

In the Lamb-Dicke regime,
the Hamiltonian describing
the carrier interaction of a pair of internal levels of a single ion with mass $M$
driven by a laser field takes the form of $H_{c}=\hbar\Omega_c(\sigma^+ e^{i\phi_c}+\sigma^- e^{-i\phi_c})$,
where $\eta=k\sqrt{\hbar/2M\nu_0}$ is the Lamb-Dicke parameter \cite{Sorensen,Leibfried} with
$k$ the wave number of the driving field and $\nu_0$ the frequency of a center-of-mass mode,
$\Omega_c$ is the Rabi frequency, $\phi_c$ is the field phase, and $\sigma^+$ ($\sigma^-$)
is the raising (lowering) ionic spin-$1/2$ operator. A Jaynes-Cummings (JC) Hamiltonian, also known as red-sideband interaction,
$H_{r}=\hbar\tilde{\Omega}_r\eta(\sigma^+ a e^{i\phi_r}+\sigma^- a^\dagger e^{-i\phi_r})$,
couples the two internal levels of the ion and one of the vibrational center-of-mass modes,
where $a$ ($a^\dagger$) is the annihilation (creation) operators of the vibrational mode.
In the blue motional sideband, also known as anti-JC (AJC) interaction, the Hamiltonian can be written as
$H_{b}=\hbar\tilde{\Omega}_b\eta(\sigma^+ a^\dagger e^{i\phi_b}+\sigma^- a e^{-i\phi_b})$,
where $\tilde{\Omega}_{r(b)}$ and $\phi_{r(b)}$ are the Rabi frequency and phase of the light field.
By applying all of these interactions simultaneously with appropriate Rabi frequencies
and relative phases, the Dirac Hamiltonian for a free particle, $H_{\rm free}=c\sigma_xp+mc^2\sigma_y$,
can be completely mapped by making the identifications $mc^2\coloneqq\hbar\Omega_c$, and $c\coloneqq2\eta\Lambda\tilde{\Omega}_1$
\cite{Lamata,Gerritsma}. Here, $p=i\hbar(a^\dagger-a)/2\Lambda$ with $\Lambda=\sqrt{\hbar/4M\nu_0}$
the size of zero-point wave packet, and
$\tilde{\Omega}_1=\tilde{\Omega}_r=\tilde{\Omega}_b$. We point out that the carrier can generate a mass term with a $\sigma_y$ Pauli matrix at lowest order, which contains the same physics as the $\sigma_z$, given that the same Clifford algebra is satisfied. Another possibility that does not employ the carrier is via a detuning in the red and blue sideband pulses, which will directly generate the $\sigma_z$ term in an appropriate interaction picture. In general, a time-dependent Rabi frequency $\Omega_c$ or detuning
will induce a simulated time-dependent mass in the Dirac system, as our protocol does. In addition, as shown in \cite{Casanova, Gerri2},
a free Dirac equation can be encoded by a single ion (ion 1), and external potentials can be implemented by a second ion (ion 2)
driven by another bichromatic light field with same vibrational mode but a different electronic transition. For example,
by imposing a laser field with appropriate phases and a time-dependent Rabi frequency $\tilde{\Omega}_2$ on the ion 2, the Hamiltonian
for the two-ion system will take the form of $H_e=c\sigma_xp+mc^2\sigma_z-e\phi_e$, where $-e$ is the electron charge, $\phi_e$ is a nonzero
electric potential, $e\phi_e\coloneqq g(t)\sigma^{(2)}_xx$ with $g(t)=\hbar\eta\tilde{\Omega}_2(t)/\Lambda$,
and $x=(a+a^\dagger)\Lambda$ is the position operator \cite{Casanova}. If ion 2 is prepared in the positive eigenstate
of Pauli operator $\sigma^{(2)}_x$, this operator could be replaced by its $+1$ eigenvalue, and this reduces
to a linear potential in the Hamiltonian $H_e$, which is in consistent with the Hamiltonian $H_u$
in Eq. (\ref{HU}),
with $\dot{\alpha}_t/c\coloneqq g(t)$. Up to a unitary transformation $U^\dagger=e^{i\alpha_{t}x/(\hbar c)}$, the
Hamiltonian $H$ of Eq.~(\ref{hamiltonian}) is found. Thus, the optimal robust quantum state engineering
protocol in Dirac dynamics can be effectively mapped by a string of two trapped ions.
Alternatively, the synthetic electric field may be implemented directly in $H$ without a second ion with a proper pulse.
Unlike the Schr\"odinger equation,
a $\pi$-carrier pulse for Dirac dynamics does not invert the population perfectly for a wave packet, see Eq. (\ref{43}),
due to the first term in $H$, a problem that may be solved by inverse-engineered
optimized pulses as the ones proposed in Sec. \ref{sec4}.
\section{Discussion and Summary} \label{sec6}
Different systems that behave according to the same model equations -with disparate interpretation
of the symbols-  simulate each other.
Often one of these systems is easier to control and manipulate.
It may also obey the model for a domain of parameters hard or impossible to implement in the other one
leading to exotic phenomena.
Dirac systems obeying the Dirac equation represent well this scenario and offer manipulation possibilities
much beyond the ones for the domain of  spin-1/2 relativistic particles.
In line with the current interest to develop quantum technologies, quantum effects beyond
the Schr\"odinger equation, as those described by a Dirac equation, are being investigated due to peculiarities
of the
spectrum, band structure, rich phase diagrams,  remarkable transport properties \cite{Dirac1,Kim,Tarruel},
and
control possibilities implied  by the coupling between internal states and momentum~\cite{Schliemann}.
This motivates the development of efficient control approaches for Dirac dynamics. The mentioned coupling
may be useful for well defined momenta,
but also limits the  controllability of internal states introducing
systematic errors for a wave packet with a nonnegligible momentum width.
We have demonstrated that inverse engineering based on invariants  of motion
provides  robust protocols for manipulating  the qubit in a 1+1
Dirac system implemented by trapped ions.  This example suggests that ``shortcuts to adiabaticity''
are a useful tool in the broad context of quantum simulations and
more generally to develop quantum technologies.
%
%
%
%
%
%
%
\section*{ACKNOWLEDGMENTS}
The authors acknowledge support from Spanish MINECO/FEDER Grants FIS2015-69983-P, FIS2015-67161-P,
Basque Government Grant IT986-16,
Ram\'on y Cajal Grant RYC-2012-11391, UPV/EHU UFI 11/55, National Natural Science
Foundation of China under Grant No. 11674033 and No.
11474026, and the Fundamental Research Funds for the
Central Universities under Grant No. 2015KJJCA01.


\begin{thebibliography}{99}

\bibitem{dirac} P. A. M. Dirac, Proc. R. Soc. A \textbf{117}, 778 (1928).

\bibitem{thaller} B. Thaller, The Dirac Equation (Springer, Berlin, 1956).

\bibitem{Lamata} L. Lamata, J. Le\'{o}n, T. Sch\"{a}tz, and E. Solano, Phys. Rev. Lett. \textbf{98}, 253005 (2007).

\bibitem{Gerritsma} R. Gerritsma, G. Kirchmair, F. Z\"{a}hringer, E. Solano, R. Blatt, and C. F. Roos, Nature (London) \textbf{463}, 68 (2010).

\bibitem{Casanova} J. Casanova, J. J. Garc\'{\i}a-Ripoll, R. Gerritsma, C. F. Roos, and E. Solano, Phys. Rev. A \textbf{82}, 020101(R) (2010).

\bibitem{Gerri2} R. Gerritsma, B. P. Lanyon, G. Kirchmair, F. Z\"ahringer, C.~Hempel, J. Casanova, J. J. Garc\'\i a-Ripoll, E. Solano, R. Blatt, and
C. F. Roos, Phys. Rev.
Lett. \textbf{106}, 060503 (2011).

\bibitem{Dirac1} T. O. Wehling, A. M. Black-Schaffer,  and A. V. Balatsky,  Adv. Phys. \textbf{63}, 1 (2014).

\bibitem{Dirac2} S. Longhi,  Optics Lett. \textbf{35}, 235 (2010).

\bibitem{Salger}T. Salger, C. Grossert, S. Kling, and M. Weitz, Phys. Rev. Lett. \textbf{107}, 240401 (2011).

\bibitem{Zhang0}D.-W. Zhang, Z.-D. Wang, and S.-L. Zhu, Front. Phys. \textbf{7}, 31 (2012).

\bibitem{Julen}J. Pedernales, R. Di Candia, D. Ballester, and E. Solano, New J. Phys. \textbf{15}, 055008 (2013).

\bibitem{deffner} S. Deffner, New J. Phys. \textbf{18}, 012001 (2016).

\bibitem{Torrontegui}  E. Torrontegui, S. Ib\'{a}\~{n}ez, S. Mart\'{\i}nez-Garaot, M. Modugno, A. del Campo,
D. Gu\'{e}ry-Odelin, A. Ruschhaupt, X. Chen, and J.~G.~Muga, Adv. At. Mol. Opt. Phys. \textbf{62}, 117 (2013).

\bibitem{Mugadirac} J. G. Muga, M. A. Sim\'{o}n and A. Tobalina, New J. Phys. \textbf{18}, 021005 (2016).

\bibitem{Ruschhaupt} A. Ruschhaupt, X. Chen, D. Alonso, and J. G. Muga, New J. Phys. \textbf{14}, 093040 (2012).

\bibitem{Lu} X. J. Lu, X. Chen, A. Ruschhaupt, D. Alonso, S. Gu\'erin, J. G.  Muga,
Phys. Rev. A \textbf{88}, 033406 (2013).

\bibitem{thaller2} B. Thaller, Advanced Visual Quantum Mechanics (Berlin: Springer, 2004).


\bibitem{ff1} S. Masuda and K. Nakamura,  Proc. R. Soc. A \textbf{466}, 1135 (2010).
\bibitem{ff2} S. Masuda  and K. Nakamura, Phys. Rev. A \textbf{84}, 043434 (2011).

\bibitem{Lewis} H. R. Lewis and W. B. Riesenfeld, J. Math. Phys. \textbf{10}, 1458 (1969).

\bibitem{Landim} R. R. Landim and I. Guedes, Phys. Rev. A \textbf{61}, 054101 (2000).

\bibitem{Castro} A. S. de Castro and A. de Souza Dutra, Phys. Rev. A \textbf{67}, 054101 (2003).

\bibitem{Khantoul} B. Khantoul and  A. Fring, Phys. Lett. A \textbf{379}, 2704 (2015).

\bibitem{zhang}Z. G. Zhang, Phys. Scr. \textbf{76}, 349 (2007).


\bibitem{Muga}  J. G. Muga, X. Chen, A. Ruschhaupt, E. Torrontegui,
and D.~Gu\'{e}ry-Odelin, J. Phys. B \textbf{42}, 241001 (2009).

\bibitem{Chen}  X. Chen, E. Torrontegui, and J. G. Muga, Phys. Rev. A \textbf{83}, 062116 (2011).


\bibitem{Daems} D. Daems, A. Ruschhaupt, D. Sugny, and S. Gu\'erin, Phys. Rev. Lett. \textbf{111}, 050404 (2013).

\bibitem{Sorensen} A. S{\o}rensen and K. M{\o}lmer, Phys. Rev. Lett. \textbf{82}, 1971 (1999).

\bibitem{Leibfried} D. Leibfried, R. Blatt, C. Monroe, and D. Wineland, Rev. Mod. Phys. \textbf{75}, 281 (2003).

\bibitem{Kim} N. Y. Kim, K. Kusudo, A. L\"offler, S. H\"ofling, A. Forchel, and Y. Yamamoto, New J. Phys. \textbf{15}, 035032 (2013).

\bibitem{Tarruel} L. Tarruell, D. Greif, T. Uehlinger, G. Jotzu, and   T. Esslinger,
  Nature \textbf{483}, 302 (2012).

\bibitem{Schliemann} J. Schliemann, D. Loss, and R. M. Westervelt,
 Phys. Rev. Lett. \textbf{94}, 206801 (2005).

\end{thebibliography}
\end{document}